\documentclass[12pt]{iopart}
\usepackage{epsfig}
\usepackage{comment}
 
\begin{document}

\title[Lunar Laser Ranging Tests of the Equivalence Principle]{Lunar Laser Ranging Tests of the Equivalence Principle}

\author{James G. Williams, Slava G. Turyshev, and Dale Boggs}

\address{Jet Propulsion Laboratory, California Institute of Technology,\\
4800 Oak Grove Drive, Pasadena, CA 91109
}
%\ead{custserv@iop.org}
\begin{abstract}
The Lunar Laser Ranging (LLR) experiment provides precise observations of the lunar orbit that contribute to a wide range of science investigations. In particular, time series of highly accurate measurements of the distance between the Earth and Moon provide unique information that determine whether, in accordance with the Equivalence Principle (EP), both of these celestial bodies are falling towards the Sun at the same rate, despite their different masses, compositions, and gravitational self-energies. Analyses of precise laser ranges to the Moon continue to provide increasingly stringent limits on any violation of the EP. Current LLR solutions give $(-0.8 \pm 1.3) \times 10^{-13}$ for any possible inequality in the ratios of the gravitational and inertial masses for the Earth and Moon, $(m_G/m_I)_{\rm E}-(m_G/m_I)_{\rm M}$. Such an accurate result allows other tests of gravitational theories. Focusing on the tests of the EP, we discuss the existing data and data analysis techniques.  The robustness of the LLR solutions is demonstrated with several different approaches to solutions.  Additional high accuracy ranges and improvements in the LLR data analysis model will further advance the research of relativistic gravity in the solar system, and will continue to provide highly accurate tests of the Equivalence Principle.

\end{abstract}

\maketitle

\section{Introduction}
\label{sec:intro}

The Equivalence Principle (EP) lies at the foundation of Einstein's  general theory of relativity; testing this fundamental assumption with the highest possible sensitivity is clearly important, particularly since it may be expected that the EP will not hold in quantum theories of gravity \cite{Turyshev:2008}. In general relativity, test objects in free-fall follow geodesics of space-time, and what we perceive as the force of gravity is instead a result of our being unable to follow those geodesics of space-time, because the mechanical resistance of matter prevents us from doing so. One can test both the validity of the EP and of the field equations that determine the geometric structure created by a mass distribution. There are two different ``flavors'' of the Principle, the weak and the strong forms of the EP that are currently tested in various experiments performed with laboratory test masses and with bodies of astronomical sizes.

In its {\it weak form} (the WEP), the Principle states that the gravitational properties of primarily strong and electro-weak interactions obey the EP. In this case the relevant test-body differences are their fractional nuclear-binding differences, their neutron-to-proton ratios, their atomic charges, etc. General relativity and other metric theories of gravity assume that the WEP is exact.  However, many gravitational theories predict observable violations of the EP at various fractional levels ranging from $10^{-13}$ down to $10^{-16}$ \cite{Damour-etal:2002}.  For instance, extensions of the Standard Model (for discussion, see \cite{Turyshev-etal:2007,Turyshev:2008}) that contain new macroscopic-range quantum fields predict quantum exchange forces that generically violate the WEP because they couple to generalized ``charges'' rather than to mass/energy alone, as in general relativity.  Therefore, even a confirmation that the WEP is not violated at some level will be exceptionally valuable, placing useful constraints on the range of possibilities in the development of a unified physical theory.

In a laboratory, precise tests of the EP can be made by comparing the free fall accelerations, $a_1$ and $a_2$, of different test bodies. When the bodies are at the same distance from the source of the gravity, the expression for the EP takes the form 
\begin{equation}
\frac{\Delta a}{a} = \frac{2(a_1- a_2)}{(a_1 + a_2)} = \left[\frac{m_G}{m_I}\right]_1-
\left[\frac{m_G}{m_I}\right]_2 = \Delta\left[\frac{m_G}{m_I}\right],
\label{eq:wep}
\end{equation}
where $m_G$ and $m_I$ are the gravitational and inertial masses of each body. 
The sensitivity of the EP test is determined by the precision of the differential acceleration measurement divided by the degree to which the test bodies differ (e.g. composition).

Currently, the most accurate results in tests of the composition-independence of acceleration rates of various masses toward the Earth were reported by ground-based laboratories \cite{Adelberger-etal:2003}. A recent experiment measured the fractional differential acceleration between Be and Ti test bodies at the level of $\Delta a/a=(0.3\pm1.8) \times 10^{-13}$ \cite{Schlamminger-etal:2008}. The accuracy of these experiments is high enough to confirm that the strong, weak, and electromagnetic interactions each contribute equally to the passive gravitational and inertial masses of the laboratory bodies.  A review of the recent laboratory tests of gravity can be found in \cite{Adelberger-etal:2003}. Significant improvements in the tests of the EP are expected from dedicated space-based experiments \cite{Turyshev:2008}. 

In its {\it strong form} (the SEP) the EP is extended to cover the gravitational properties resulting from gravitational energy itself, thus involving an assumption about the non-linear property of gravitation \cite{Williams-etal:2009}. In the SEP case, the relevant test body differences are the fractional contributions to their masses by gravitational self-energy. Although general relativity assumes that the SEP is exact, many modern theories of gravity typically violate the SEP by including new fields of matter, notably scalar fields \cite{Damour-Nordtvedt:1993,Damour-Polyakov:1994}. The ratio of the gravitational-to-inertial masses for a body can be parameterized as 
\begin{equation}
\left[\frac{m_G}{m_I}\right] = 1+ \eta \frac{U}{mc^2},
\label{eq:eta}
\end{equation}
where $m$ is the mass of a body, $U$ is the body's gravitational self-energy ($U< 0$), $mc^2$ is its total mass-energy, and $\eta$ is a dimensionless constant for SEP violation ($\eta =0$ in general relativity, see discussion in \cite{Williams-etal:2009}). Because of the extreme weakness of gravity, a test of the SEP requires bodies of astronomical sizes.

At present, the Earth-Moon-Sun system provides the best solar system arena for testing the SEP. Time series of the highly accurate measurements of the distance between the Earth and Moon provide unique information used to determine whether, in accordance with the EP, both of these celestial bodies are falling towards the Sun at the same rate, despite their different masses, compositions, and gravitational self-energies. 

This paper focuses on the tests of the EP with LLR. To that extent, Section~\ref{sec:2} discusses the LLR history, experimental technique, and the current state of the effort.  Section ~\ref{sec:3} discussed LLR range data and distribution. Section~\ref{sec:4} focuses on the precision tests of the EP with LLR data and discusses the results obtained. Section~\ref{sec:5} discusses future prospects. In Section \ref{sec:conc} we conclude with a summary and outlook.

\section{LLR Technique}
\label{sec:2}

Over 40 years since their initiation, analyses of precision laser ranges to the Moon continue to provide increasingly stringent limits on any violation of the EP.

Each LLR measurement is the round-trip travel time of a laser pulse between an observatory on the Earth and one of the five corner-cube retro-reflector (CCR) arrays on the moon.  To range the moon, the observatories fire a short laser pulse toward the target array.  The lasers currently used for ranging operate at 20 Hz, with a pulse width of about 200 psec; each pulse contains $\sim10^{18}$ photons. Under favorable observing conditions a single reflected photon is detected every few seconds for most LLR stations and in less than one second for Apache Point \cite{Murphy-etal:2007}. Such a low return rate is due to the huge attenuation encountered during the round-trip of the pulse. The outgoing narrow laser beam must be accurately pointed at the target since the beam's angular spread is typically a few arcsec; the spot size on the moon is a few km across. The amount of energy falling on the CCR depends inversely on that spot area. 

The returning pulse illuminates an area around the observatory that is a few tens of kilometers in diameter ($\sim$10 km for the 532~nm green light and Apollo CCRs).  The observatory has a sensitive detector which records single photon arrivals.  The power received by the telescope depends directly on the telescope's collecting area and inversely on the returning spot area.  Velocity-caused aberration of the returning beam is ($\sim$1 arcsec.  

\subsection{Equivalence Principle and the Earth-Moon system}

The Jet Propulsion Laboratory's (JPL) program which integrates the orbits of the Moon and planets considers accelerations due to Newtonian, geophysical and post-Newtonian effects. The dynamics of the three-body Sun-Earth-Moon system in the solar system barycentric inertial frame provides the main LLR sensitivity for a possible violation of the equivalence principle. In this frame, the quasi-Newtonian acceleration of the Moon with respect to the Earth, ${\bf a} = {\bf a}_M - {\bf a}_E$, is calculated to be:
{}
\begin{equation}
{\bf a} = -\mu^* \frac{{\bf r}_{\rm EM}}{r^3_{\rm EM}} - \left[\frac{m_G}{m_I}\right]_{\rm M} \mu_{\rm S} \frac{{\bf r}_{\rm SM}}{r^3_{\rm SM}} + \left[\frac{m_G}{m_I}\right]_{\rm E}\mu_{\rm S}\frac{{\bf r}_{\rm SE}}{r^3_{\rm SE}},
\end{equation}
where $\mu^* = \mu_{\rm E} (m_G/m_I)_{\rm M} + \mu_{\rm M} (m_G/m_I)_{\rm E}$ and $\mu_k = G m_k$.  The first term on the right-hand side of the equation above, is the acceleration between the Earth and Moon with the remaining pair being the tidal acceleration expression due to the solar gravity. The above acceleration is useful for either the weak or strong forms of the EP. 

Rearranging the above equation emphasizes the EP terms:
{}
\begin{eqnarray}
{\bf a} = -\mu^* \frac{{\bf r}_{\rm EM}}{r^3_{\rm EM}} &+&\mu_{\rm S}\left[\frac{{\bf r}_{\rm SE}}{r^3_{\rm SE}} - \frac{{\bf r}_{\rm SM}}{r^3_{\rm SM}}\right] + \nonumber\\
&+&\mu_{\rm S} \left[\bigg(\left[\frac{m_G}{m_I}\right]_{\rm E}-1\bigg) \frac{{\bf r}_{\rm SE}}{r^3_{\rm SE}} -\bigg(\left[\frac{m_G}{m_I}\right]_{\rm M}-1\bigg) \frac{{\bf r}_{\rm SM}}{r^3_{\rm SM}}\right].
\end{eqnarray}
The presence of the EP parameters for gravitational-to-inertial mass ratios in $\mu^*$ modifies Kepler's third law to $n^2 a^3 = \mu^*$ for the relation between semimajor axis $a$ and mean motion $n$ in the elliptical orbit approximation. This term is notable, but in the LLR solutions $\mu_{\rm E} +\mu_{\rm M}$ is a solution parameter with uncertainty, so this term does not provide a sensitive test of the EP, though its effect is implicit in the LLR solutions. The second term on the right-hand side with the differential acceleration toward the Sun is the Newtonian tidal acceleration. The third term involving the $m_G/m_I$ ratios for two bodies gives the main sensitivity of the LLR test of the EP. Since the distance to the Sun is $\sim$390 times the distance between the Earth and Moon, the last term is approximately the difference of any EP violation of the two bodies times the Sun's acceleration of the Earth-Moon center of mass. 

Treating the EP related tidal term as a perturbation \cite{Nordtvedt:1968c} found a polarization of the Moon's orbit in the direction of the Sun with a radial perturbation
{}
\begin{equation} 
\Delta r = S [(m_G/m_I)_{\rm E} - (m_G/m_I)_{\rm M}] \cos D,
\label{eq:delta-r}
\end{equation} 
where $S$ is a scaling factor of about $-2.9 \times 10^{13}$ mm (see \cite{Nordtvedt:1995,Damour-Vokrouhlicky:1996a}).  

The Earth and Moon are large enough to have significant gravitational self-energies and a lunar test of the EP was proposed by Nordtvedt \cite{Nordtvedt:1968c}.  Both bodies have differences in their compositions and self-energies and the Sun provides the external gravitational acceleration. For the SEP effect on the Moon's position with respect to the Earth it is the difference of the two 
%accelerations and 
self-energy values which is of interest \cite{Williams-etal:1996}:
\begin{equation}
\left(\frac{U}{mc^2}\right)_{\rm E} - \left(\frac{U}{mc^2}\right)_{\rm M} = -4.45 \times 10^{-10}.
\label{eq:self-E}
\end{equation}

For the SEP, combining Eqs.~(\ref{eq:eta}) and (\ref{eq:delta-r}) yields 
\begin{eqnarray}
\Delta r &=& S \eta \left[\left(U/mc^2\right)_{\rm E} - \left(U/mc^2\right)_{\rm M}\right] \cos D, 
\label{eq:S}\\
\Delta r &=& C_0 \eta \cos D. \label{eq:C0}
\end{eqnarray}

Applying the difference in numerical values for self-energy for the Earth and Moon gives a value of $C_0$ of about 13 m (see \cite{Nordtvedt-etal:1995,Damour-Vokrouhlicky:1996a}). In general relativity $\eta$ = 0. A unit value for $\eta$ would produce a displacement of the lunar orbit about the Earth, causing a 13 m monthly range modulation.  (See Sec.~\ref{sec:4} and Table~\ref{tab:3} for a comparison of the theoretical values of $S$ and $C_0$ with numerical results.) 

In essence, LLR tests of the EP compare the free-fall accelerations of the Earth and Moon toward the Sun. Lunar laser-ranging measures the time-of-flight of a laser pulse fired from an observatory on the Earth, bounced off of a retro-reflector on the Moon, and returned to the observatory (see \cite{Dickey-etal:1994,Williams-etal:2009}). If the EP is violated, the lunar orbit will be displaced along the Earth-Sun line, producing a range signature having a 29.53 day synodic period (different from the lunar orbit period of 27 days). For a review of history and present capabilities of LLR to test the EP, see \cite{Williams-etal:2009}. 

\subsection{LLR Normal Points, Model and Science Outcome}

A normal point results from a statistical combination of the observed transit times of several individual photons arriving at the observing detector within a relatively short time span, typically minutes to tens of minutes \cite{Samain-etal:1998,Murphy-etal:2008}. The resulting ``range'' normal point is the round trip flight time for a particular firing time. In addition to range and time, supporting information includes uncertainty, signal to noise ratio, number of photons, atmospheric pressure, temperature, and humidity.

The existing model formulation at JPL and its computational realization in computer code, is the product of many years of effort \cite{Dickey-etal:1994,Williams-etal:1996}.  The successful analysis of LLR data requires attention to geophysical and rotational effects for the Earth and the Moon in addition to orbital effects. The main features of the current model are summarized in \cite{Williams-etal:1996,Williams-etal:2004b,Standish-Williams:2012}.  For a general review of LLR see \cite{Dickey-etal:1994}. Refs.~\cite{Williams-etal:2004a,Williams-etal:2004b,Williams-etal:2006,
Williams-etal:2009} have the most recent results for gravitational physics.

\section{Range Data and Distribution}
\label{sec:3}

The solutions presented here use 17,580 laser ranges from March 1970 to July 2011. The ranging stations include the McDonald Observatory in Texas (37.7\% of the total observations), the Observatoire de la C\^ote d'Azur (OCA) in France (52.9\%), the Haleakala Observatory in Hawaii (3.9\%), and the APOLLO facility at the Apache Point Observatory in New Mexico (5.4\%). The OCA station is described by \cite{Samain-etal:1998}. The APOLLO effort is described in the article by \cite{Murphy-etal:2012} in this issue. The Matera station in Italy has demonstrated lunar capability with 0.1\% of the total observations, but does not range the Moon regularly. There are five retro-reflectors on the Moon at the Apollo 11, 14 and 15 sites and on the Lunokhod 1 and 2 rovers. They contribute 10.1\%, 10.0\%, 76.9\%, 0.2\%, and 2.8\% of the total observations, respectively. The Apollo 15 retro-reflector gives the strongest signal because it is largest. The Lunokhod 1 location was only recently determined with sufficient accuracy for ranging after being photographed by the Lunar Reconnaissance Orbiter \cite{Murphy-etal:2011}. It gives a stronger return signal than the Lunokhod 2 retro-reflector and will be useful in the future. LLR data are archived by the ILRS \cite{Pearlman-etal:2002}. More information on the ranging stations and technique is in \cite{Mueller-etal:2012}.

\begin{comment}
%************
\begin{figure}[!t]
 \begin{center}
\noindent    
\epsfig{figure=Annual-RMS-1970-2011.eps,width=90mm}%,height=90mm}
\end{center}
\vskip -10pt 
  \caption{Annual rms residuals of LLR data from 1970 to 2011.  
 \label{fig:1}}
\end{figure} 
%**************

%************
\begin{figure}[!t]
 \begin{center}
\noindent    
\epsfig{figure=D-Hist-All-1970-2011.eps,width=90mm}%,height=90mm}
\end{center}
\vskip -10pt 
  \caption{A histogram of the distribution of LLR observations vs. elongation angle $D$ for all of the data.  
 \label{fig:1}}
\end{figure} 
%**************
\end{comment}

%**************
\begin{figure}[t]
\begin{minipage}[b]{.49\linewidth}
\includegraphics[width=0.97\linewidth]{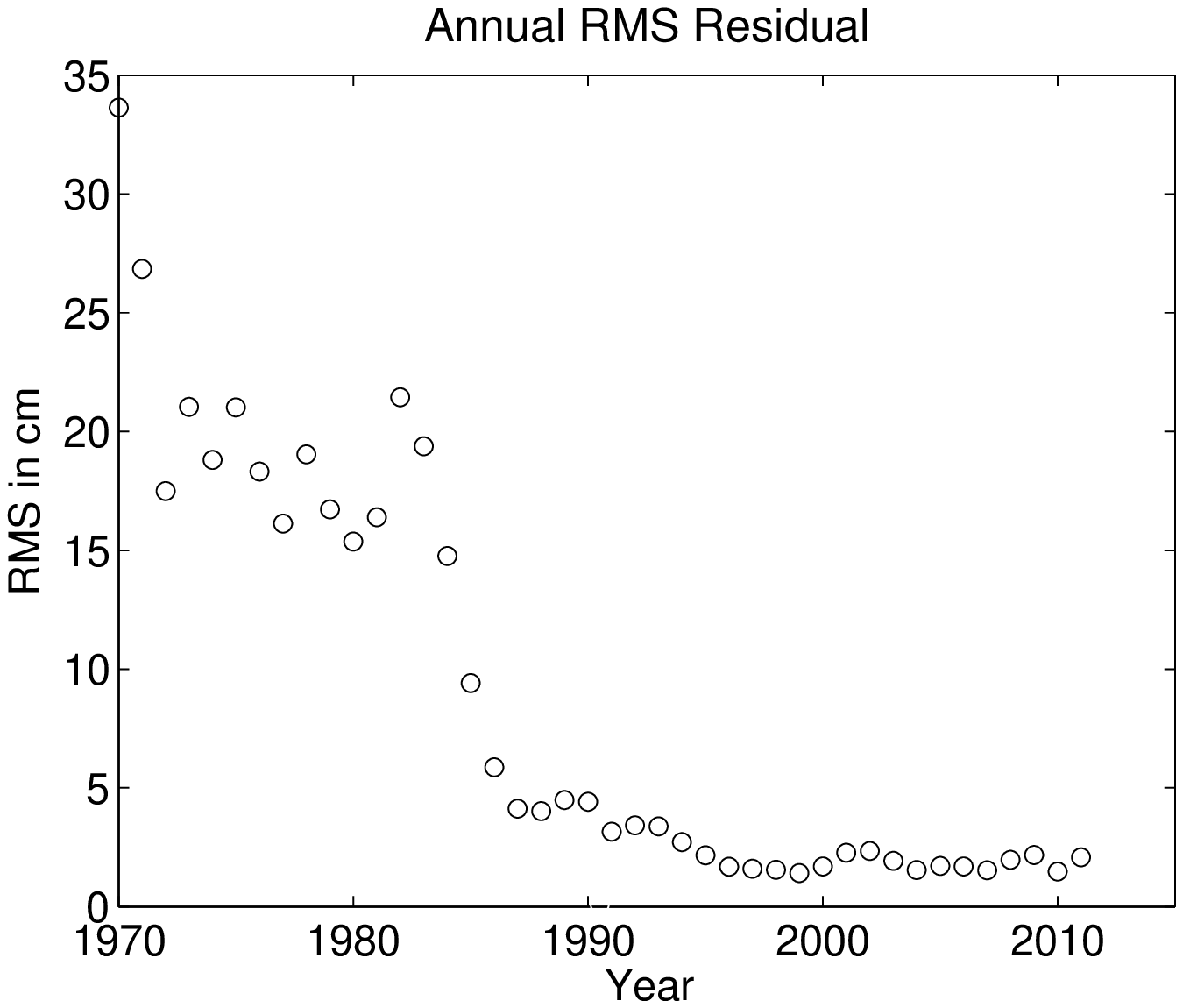}
\end{minipage}
\begin{minipage}[b]{.49\linewidth}
\includegraphics[width=0.97\linewidth]{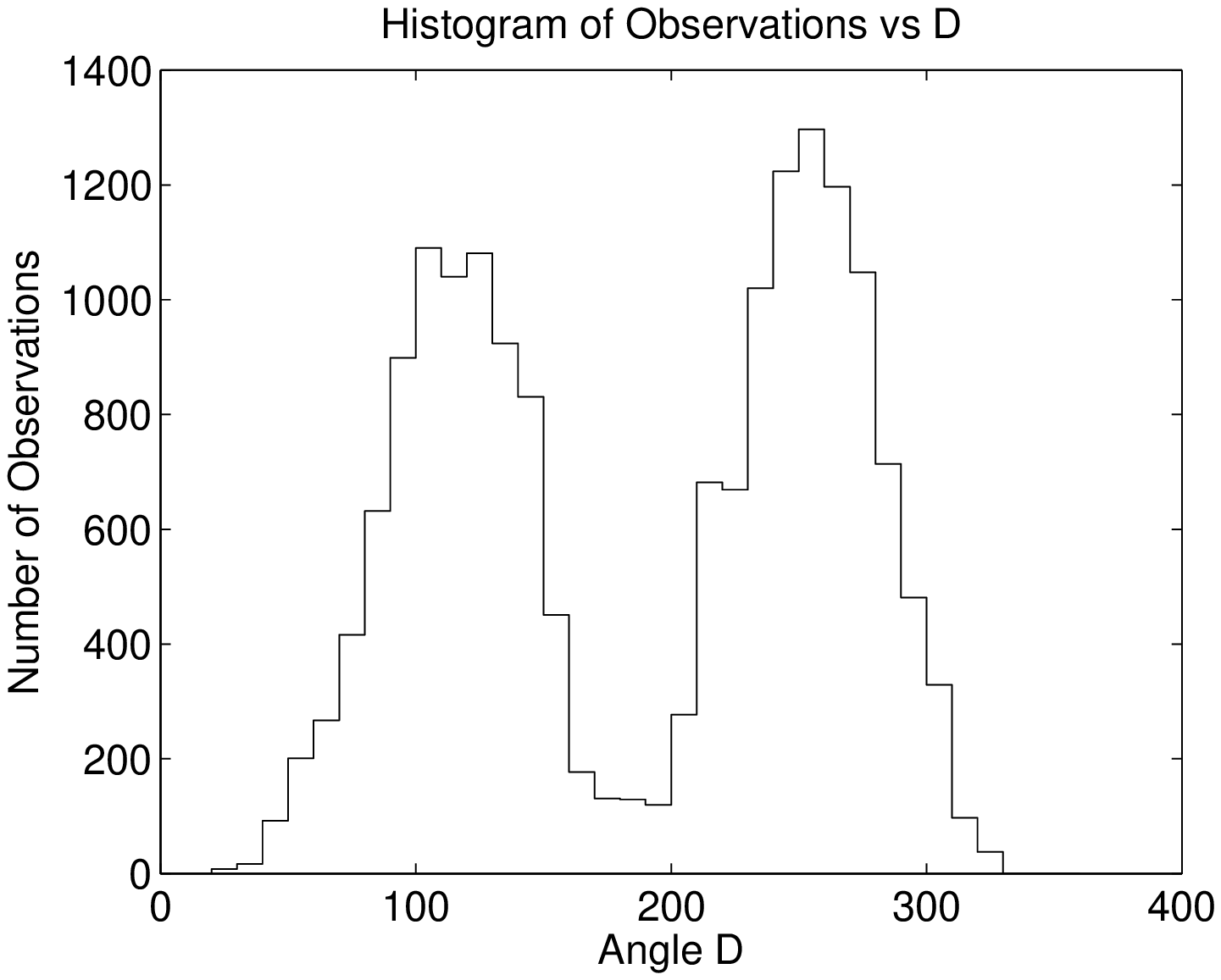}
\end{minipage}
\vskip -2pt
\caption{\label{fig:1} (a) Annual rms residuals of LLR data from 1970 to 2011.  (b) A histogram of the distribution of LLR observations vs. elongation angle $D$ for all of the data.}
\end{figure}
%**************

Figure~\ref{fig:1}a shows the post-fit weighted rms residuals for each year from 1970 to 2011. The prominent decreases correspond to equipment improvements. Ranges from 2005-2011 are fit with a 1.8 cm weighted rms residual. The weights depend on the range accuracy provided by each station and the analysis software's ability to fit the ranges. For the equivalence principle test, LLR analyses are looking for a $\cos D$ signature so the distribution with $D$, the mean lunar elongation from the Sun, is important. Figure~\ref{fig:1}b shows a histogram of the distribution of LLR observations vs. elongation angle $D$ for all of the data. The bright Sun prevents observations near new Moon and there are few observations near full Moon. Most of the observations near full Moon are less accurate early observations. Alone among the modern more accurate stations, the APOLLO station provides some ranges near full Moon \cite{Murphy-etal:2010}. The nonuniform distribution with $D$ weakens the LLR equivalence principle test. 

\section{Data Analysis}
\label{sec:4}

This section presents analysis of the lunar laser ranging data to test the equivalence principle.  To check consistency, more than one solution is presented.  Solutions are made with two different equivalence principle parameters and several different choices for other parameters.  The solutions are compared and discussed, and residuals after fits, the post-fit residuals, are examined for systematics.  

An EP violation can be solved for in two ways.  The first is a parameter for $(m_G/m_I)_{\rm E} - (m_G/m_I)_{\rm M}$ with a dynamical partial derivative generated from numerical integration. The second solves for a coefficient of $\cos D$ in the lunar range, a one-term representation. The latter approach was used in early papers, but the more sophisticated dynamical parameter is used in subsequent publications \cite{Williams-etal:2004b,Turyshev-Williams:2007}. Both approaches are exercised here to investigate consistency.  

Six equivalence principle solutions are presented in Table~\ref{tab:1}.  Each of these cases solves for a standard set of Newtonian parameters in addition to one or more equivalence principle parameters. Among the Newtonian solution parameters are $Gm_{\rm Earth+Moon}$, lunar orbit parameters including semimajor axis, Moon-centered retroreflector coordinates, geocentric ranging station coordinates, lunar tidal displacement Love number $h_2$, and parameters associated with the lunar orientation (physical librations) and physical properties. The interior model for the physical librations includes a fluid core, but the size of that core remains uncertain. Consequently, two different fluid core moments were used, $3 \times 10^{-4}$ and $7 \times 10^{-4}$ compared to the total lunar moment. These cases have designations starting with EP3 and EP7, respectively. The annual nutation of the Earth's pole direction in space involves four parameters, the in-phase and out-of-phase components of two nutation angles. We normally solve for these four nutation parameters in LLR solutions because they are sensitive to the flattening of the Earth's core/mantle boundary, and this was done for cases EP31 and EP71. The LLR uncertainties for the two in-phase components are about 0.23 milliarcseconds (mas). The out-of-phase components have 0.17 mas LLR uncertainties. In these two solutions the equivalence principle correlates 0.90 (ignoring signs) with the two in-phase annual nutation components, which weakens the EP uncertainty. Consequently, the remaining solutions fix the four annual nutation components at the IAU 2000A nutation model values given in the 2003 IERS conventions \cite{McCarthy-Petit:2003}. This is justified because the nutation model is compatible with accurate VLBI solutions for terrestrial nutations at the level required for the study here. When the annual nutation components are fixed the EP uncertainty improves by more than a factor of two as solutions EP32 and EP72 show. While the first four cases in the table solve for $(m_G/m_I)_{\rm E} - (m_G/m_I)_{\rm M}$ using the numerically integrated partial derivative, the EP73 case solves for coefficients of $\cos D$ and $\sin D$ in range as an alternative. Only the $\cos D$ coefficient behaves like the EP. The $\sin D$ component shows nothing above the noise level. The last case, EP74, provides a check of the Eq.~(\ref{eq:S}) scaling factor $S$ that relates $(m_G/m_I)_{\rm E} - (m_G/m_I)_{\rm M}$ violations to the $\cos D$ coefficient, as discussed below. The column labeled ``converted $\cos D$ coef'' converts $(m_G/m_I)_{\rm E} - (m_G/m_I)_{\rm M}$ to the coefficient of $\cos D$ in radial distance using the factor $S = -2.94 \times 10^{13}$ mm in Eq.~(\ref{eq:S}).  The sum of the converted and solution $\cos D$ coefficients is $0.69 \pm 3.80$ mm, using the 0.9790 correlation, in agreement with the estimates of the single EP parameter cases.  

%====================================
\begin{table}[t!]
%\begin{center}
\caption{Six solutions for the equivalence principle.}  %\vskip 0pt
{\begin{tabular}{cccccc} \hline\hline
{Solution} & {Ann Nut} & $(m_G/m_I)_{\rm E}-$
&{\small Converted}  &  $\cos D$ {{\rm coef}} &  $\sin D$ {{\rm coef}} \\
ID &   & $ ({m_G}/{m_I})_{\rm M}$ & $\cos D$ {\small {\rm coef}} & & \\\hline
Units& & $ \times \,10^{-13}$ &mm& mm & mm \\\hline
EP31 & solved & $1.03\pm3.39$ & $-3.03\pm9.97$ &&\\
EP71 & solved & $0.65\pm3.35$ & $-1.91\pm9.85$ &&\\
EP32 & fixed  & $0.42\pm1.32$ & $-1.24\pm3.88$ &&\\
EP72 & fixed  & $0.03\pm1.28$ & $-0.08\pm3.75$ &&\\
EP73 & fixed  & && $0.90\pm 3.78$ & $2.43\pm2.14$ \\
EP74 & fixed  & $2.80 \pm 6.32$ & $-8.24\pm 18.57$ & $8.93\pm 18.50$ & $2.35\pm2.14$ \\
\hline\hline
\end{tabular} \label{tab:1}}
%\end{center} 
\end{table}
%========================================================== 

Solar radiation pressure, like the acceleration from an equivalence principle violation, is aligned with the direction away from the Sun and it produces a perturbation with the 29.53 d synodic period.  This force on the Earth and Moon must be considered for the most accurate tests of the equivalence principle.  This acceleration is not currently modeled in the JPL software.  Here we rely on the analysis of Vokrouhlicky \cite{Vokrouhlicky:1997} who considered incident and reflected radiation for both bodies plus thermal radiation from the Moon.  He finds a solar radiation perturbation of $-3.65\pm0.08$ mm $\cos D$ in the radial coordinate. 

Thermal expansion of the corner cube mounts is another effect worth consideration. The peak-to-peak variation of surface temperature at low latitudes on the Moon is nearly 300$^\circ$.  The lunar ``day'' is 29.53 days long.  This is the same period as the main equivalence principle term so a systematic effect from thermal expansion is expected.  The phase of the thermal cycle depends on the retro-reflector longitude. If the Apollo arrays share the same temperature variations as the surface, then the total vertical variation of thermal expansion will be 1 to 2 mm and the $\cos D$-like amplitude would be about $2/3$ that ($2/p$ for a square wave).  The Apollo arrays make up 97\% of the data so the range effect could be as much as 1.0 mm $\cos D$, but will be less if the array temperature variations are less than the lunar surface variations. 

For the sum of the thermal and solar radiation pressure effect we use $-3.0\pm 0.5$~mm $\cos D$. The correction to Table~\ref{tab:1} has opposite sign; this correction is used to construct Table~\ref{tab:2} For the EP74 case the partitioning of the correction for the two highly correlated EP parameters is not clear and that case is not given in Table~\ref{tab:2}. However, the sum of those two EP parameters would give $3.69\pm3.83$ mm $\cos D$, equivalent to $(-1.26\pm1.30)\times 10^{-13}$ for $(m_G/m_I)_{\rm E} - (m_G/m_I)_{\rm M}$. 

%====================================
\begin{table}[t!]
%\begin{center}
\caption{Five solutions for the equivalence principle corrected for solar radiation pressure and thermal expansion.}  %\vskip 0pt
{\begin{tabular}{cccccc} \hline\hline
{Solution} & {Ann Nut} & $(m_G/m_I)_{\rm E}-$
&{\small Converted}  &  $\cos D$ {{\rm coef}} &  $\sin D$ {{\rm coef}} \\
ID &   & $ ({m_G}/{m_I})_{\rm M}$ & $\cos D$ {\small {\rm coef}} & & \\\hline
Units& & $ \times \,10^{-13}$ &mm& mm & mm \\\hline
EP31 & solved & $~~0.01\pm3.40$ & $-0.03\pm9.98$ &&\\
EP71 & solved & $-0.37\pm3.35$ & $~~1.09\pm9.86$ &&\\
EP32 & fixed  & $-0.60\pm1.33$ & $~~1.76\pm3.91$ &&\\
EP72 & fixed  & $-0.99\pm1.29$ & $~~2.92\pm3.78$ &&\\
EP73 & fixed  & && $3.90\pm 3.81$ & $2.43\pm2.14$ \\
\hline\hline
\end{tabular} \label{tab:2}}
%\end{center} 
\end{table}
%========================================================== 

The EP solution parameters in Tables~\ref{tab:1} and \ref{tab:2} are within their uncertainties for all cases except the EP73 case in Table~\ref{tab:2}, and that value is just slightly larger. The EP73 coefficient of $\cos D$ and the sum of terms for EP74 are about 1 mm larger than the value from the converted $(m_G/m_I)_{\rm E} - (m_G/m_I)_{\rm M}$ parameter of the EP72 case. The uncertainties are compatible. Solutions with the two lunar interior models differ by about 1 mm with the smaller lunar core giving smaller $\cos D$ coefficients. Solutions with different EP parameters are compatible. There is no evidence for a violation of the EP from any of the solutions. 

The EP32 and EP72 solutions are the preferred cases since they use numerically integrated partial derivatives for $(m_G/m_I)_{\rm E} - (m_G/m_I)_{\rm M}$. The average of those two corrected solutions is $(m_G/m_I)_{\rm E} - (m_G/m_I)_{\rm M} = (-0.8\pm1.3)\times 10^{-13}$. Some larger correlations from the EP72 solution follow.  The correlation of $(m_G/m_I)_{\rm E} - (m_G/m_I)_{\rm M}$ with both $Gm_{\rm Earth+Moon}$ and osculating semimajor axis, at the 1969 epoch of the integration, is 0.31.  $Gm$ and mean semimajor axis are tightly connected through Kepler's third law since the mean motion is very well determined.  The product of mean semimajor axis and mean eccentricity is well determined and the correlation of $(m_G/m_I)_{\rm E} - (m_G/m_I)_{\rm M}$ with osculating eccentricity is 0.29. The displacement Love number $h_2$ is correlated 0.28 with $(m_G/m_I)_{\rm E} - (m_G/m_I)_{\rm M}$. All of these correlations are modest. 

The difference in uncertainty between the $\sin D$ and $\cos D$ components of both the EP73 and EP74 solutions is due to the nonuniform distribution of observations with respect to $D$, as illustrated in Fig.~\ref{fig:1}b. The $\sin D$ coefficient is well determined from observations near first and last quarter Moon where $\sin D$ has its extreme values of $+1$ and $-1$, respectively. The $\cos D$ coefficient is weakened by the absence of data close to new Moon and its scarcity near full Moon, where $\cos D$ has its extreme values.  

The EP74 case, solving for $(m_G/m_I)_{\rm E} - (m_G/m_I)_{\rm M}$ along with $\cos D$ and $\sin D$ coefficients, is instructive.  The correlation between the $(m_G/m_I)_{\rm E} - (m_G/m_I)_{\rm M}$ and $\cos D$ parameters is 0.9790 so the two quantities are nearly equivalent, as expected.  The uncertainty for each of the two $\cos D$ coefficients increases by a factor of five in the joint solution, but the sum is $0.69\pm3.80$ mm, which corrects to $3.69\pm3.83$ mm, in agreement with the other solutions. The solution is not singular, so the solution has some ability to distinguish between the two formulations.  The integrated partial derivative implicitly includes terms at frequencies other than the $D$ argument and it will also have some sensitivity to the EP influence on lunar orbital longitude.  The EP perturbation on lunar orbital longitude is about twice the size of the radial component and it depends on $\sin D$.  The ratio of Earth radius to lunar semimajor axis is $R_{\rm E}/a ˜\sim 1/60.3$, the parallax is $\approx1^\circ$, so the $\sin D$ longitude component projects into range at the few percent level.  

%====================================
\begin{table}[t!]
%\begin{center}
\caption{Theoretical and numerical results for factors $S$ and $C_0$.}  %\vskip 0pt
{\begin{tabular}{rccc} \hline\hline
{Source} & {Type} & $S$ & $C_0$ \\\hline
Nordtvedt (1995) \cite{Nordtvedt:1995} & 
           analytical & $-2.9\times 10^{13}$~mm & 12.90 m\\
Damour \& Vokrouhlicky (1996) \cite{Damour-Vokrouhlicky:1996a} & 
           analytical & $-2.9427\times10^{13}$~mm & 13.10 m\\
Nordtvedt \& Vokrouhlicky (1997) \cite{Nordtvedt-Vokrouhlicky:1997} & 
           analytical  & $-2.943 \times 10^{13}$~mm & 13.10 m\\
This paper & numerical  & $-2.992 \times 10^{13}$~mm & 13.31 m \\
\hline\hline
\end{tabular} \label{tab:3}}
%\end{center} 
\end{table}
%========================================================== 

In the radial $\cos D$ expressions the coefficients of the EP equations (\ref{eq:delta-r}) and (\ref{eq:S}) have an $S$ factor, while the strong EP $C_0$ factor of Eq.~(\ref{eq:C0}) is the product of $S$ and the difference in Earth and Moon self-energies. Three theoretical values for $S$ are given in Table~\ref{tab:3} while the $C_0$ values are based on the self-energy difference given in Eq.~(\ref{eq:self-E}). The uncertainties in the EP74 solution can be used to check the theoretical computation of the coefficients $S$ and $C_0$.  Given the high correlation between the $(m_G/m_I)_{\rm E} - (m_G/m_I)_{\rm M}$ and $\cos D$ parameters, a first approximation of $S=-2.93\times10^{13}$ mm is given by the ratio of uncertainties, and our knowledge that $S$ must be negative.  A more sophisticated estimate of $S=-2.992\times10^{13}$ mm comes from computing the slope of the axis of the uncertainty ellipse for the two parameters $C_0$ and $(m_G/m_I)_{\rm E}- (m_G/m_I)_{\rm M}$.  Using Eq.~(\ref{eq:self-E}) for the difference in self energies of the Earth and Moon, the two preceding $S$ values give $\Delta r=13.0$~m $\eta \cos D$ and $\Delta r=13.3$~m~$\eta \cos D$, respectively. As explained in the preceding paragraph, the LLR solution for $(m_G/m_I)_{\rm E} - (m_G/m_I)_{\rm M}$ involves more than a $\cos D$ radial expression so we expect a few percent uncertainty in the numerical values. 

Our earlier paper \cite{Williams-etal:2009}  discusses the importance to the LLR equivalence principle solutions of parameters that affect the mean distance of the lunar retroreflectors including $Gm_{\rm Earth+Moon}$ and lunar displacement Love number $h_2$. Both were solution parameters in all cases of Tables~\ref{tab:1} and \ref{tab:2}. 

Six solutions presented in this analysis section have tested the equivalence principle. They do not show evidence for a significant violation of the equivalence principle.  

\section{Future Prospects}
\label{sec:5}

In the fits, the data is weighted using both uncertainties assigned by the ranging stations and a contribution from the analysis software. The latter depends on the post-fit rms residuals by station and time. Modern high-quality ranges have smaller instrumental uncertainties than the analysis software's ability to fit over decades-long time spans. There have been software improvements since \cite{Williams-etal:2004b,Williams-etal:2009} but further improvements are needed. Most obvious in the highest accuracy residuals are systematic signatures that are different for different retro-reflectors. This indicates that there are unmodeled variations in the physical librations. The structure and geophysical properties of the deep lunar interior are imperfectly known. Further software improvements will advance the LLR equivalence principle tests. These upgrades are planned. 

The accuracy of the annual nutation will become more of a concern for future improved LLR tests of the equivalence principle. Four of the solutions given in this paper fix the values of the four annual nutation coefficients and the EP uncertainties do not account for uncertainty in the annual nutation. It is estimated that 0.05 milliarcsec uncertainties in the two in-phase annual nutation coefficients would increase the LLR EP uncertainties by about 10\%. 

Future ranging devices on the Moon might take two forms, namely passive retro-reflectors and active transponders. The advantages of passive retro-reflector arrays are their long life and simplicity. The disadvantages are the weak returned signal and the spread of the reflected pulse arising from lunar orientation with respect to the direction toward the Earth, which can alter the incoming direction up to 10$^\circ$ with respect to the retro-reflector face. A single large corner cube would not have this problem \cite{Currie-etal:2010}. Additional ranging devices on the Moon would have benefits for fundamental physics, lunar science, control networks for surface mapping, and navigation \cite{Turyshev-Williams:2007,Turyshev-etal:2007,Turyshev:2008}. 

\section{Summary}
\label{sec:conc}

If the ratio of gravitational mass to inertial mass is not invariant, then there would be profound consequences for gravitation.  Such a violation of the EP would affect how bodies move under the influence of gravity.  The EP is not violated in general theory of relativity, but violations are expected for many alternative theories of gravitation.  Consequently, tests of the EP are important to the search for a new theory of gravity.  

The equivalence principle (EP) is considered in its two forms; the weak equivalence principle (WEP) is sensitive to composition while the strong equivalence principle (SEP) considers possible sensitivity to the gravitational energy of a body.  The main sensitivity of the lunar orbit to the equivalence principle comes from the acceleration of the Earth and Moon by the Sun.  Any difference in those accelerations due to a failure of the equivalence principle causes an anomalous term in the lunar range with the 29.53 d synodic period.  The amplitude would be proportional to the difference in the gravitational to inertial mass ratios for Earth and Moon.  Thus, lunar laser ranging is sensitive to a failure of the equivalence principle due to either the WEP or the SEP.  In the case of the SEP, any violation of the equivalence principle can be related to a linear combination of the parameterized post-Newtonian parameters $\beta$ and $\gamma$.  

Section~\ref{sec:3} discusses the data and its distribution with respect to the new-full-new Moon cycle.  The evolution of the data from decimeter to centimeter quality fits is illustrated. For the LLR equivalence principle tests, selection with phase of the Moon is an important consideration.  

An accurate model and analysis effort is needed to exploit the lunar laser range data to its full capability.  The model is the basis for the computer code that processes the range data.  Further modeling efforts will be necessary to fully benefit from range data of millimeter quality.  Two small effects for future modeling are thermal expansion and solar radiation pressure.  An improved model of the deep lunar interior is needed for lunar orientation. 

Solutions testing the EP are given in Sec.~\ref{sec:4}.  Several approaches to the solutions are used as checks.  The equivalence principle solution parameter can be either a ratio of gravitational to inertial masses or as a coefficient of a synodic term in the range equation.  The results are compatible in value and uncertainty. In all, six equivalence principle solutions are presented in Table~\ref{tab:1} and five are corrected and carried forward into Table~\ref{tab:2}. The analysis of the LLR data does not show significant evidence for a violation of the equivalence principle compared to its uncertainty.  The final result for $(m_G/m_I)_{\rm E} - (m_G/m_I)_{\rm M}$ is $(-0.8\pm1.3)\times 10^{-13}$.  

The accuracy of the LLR test of the EP can be improved in the future. Adding new LLR data will help. The analysis software needs to be improved so that the most accurate data, including the very accurate APOLLO data, can be more fully exploited. Future 
landers on the Moon should carry new retro-reflectors with improved design. 

The lunar laser ranging results for the EP are consistent with the assumptions of Einstein's general theory of relativity.  It is remarkable that general relativity has survived a century of testing.  Each new significant improvement in accuracy is unknown territory and that is reason for future tests of the equivalence principle.

\subsection*{Acknowledgments}
The research described in this paper was carried out at the Jet Propulsion Laboratory, California Institute of Technology, under a contract with the National Aeronautics and Space Administration. Government sponsorship is acknowledged.

\end{document}